\documentclass[preprint2,twocolappendix]{aastex6}
\usepackage{color}
\usepackage{epstopdf}
\usepackage[]{natbib}

\newcommand{\kms}{km\,s$^{-1}$}
\shorttitle{SISMA, an HARPS archive of variable and active stars}
\shortauthors{Rainer et al.}
\begin{document}
\title{
The {\it SpaceInn}--SISMA database: characterization of a large sample of variable and active stars by means of HARPS spectra
}

\author{M. Rainer, E. Poretti, A. Mist{\`{o}}, R. Panzera\altaffilmark{1}}
\affil{INAF-Osservatorio Astronomico di Brera, via E. Bianchi 46, 23807 Merate, Italy}
\email{monica.rainer@brera.inaf.it}
\author{M. Molinaro, F. Cepparo\altaffilmark{2}}
\affil{INAF-Osservatorio Astronomico di Trieste, via Tiepolo 11, 34143, Trieste, Italy}
\author{M. Roth\altaffilmark{3}}
\affil{Kiepenheuer-Institut f{\"u}r Sonnenphysik, Sch{\"o}neckstr. 6, 79104 Freiburg, Germany}
\author{E. Michel\altaffilmark{4}}
\affil{Observatoire de Paris, LESIA, UMR8109, Universit\`{e} Pierre et Marie Curie, Universit\`{e} Paris Diderot, PSL, 5 pl. J. Jassen, 92195 Meudon, France}
\and
\author{M.J.P.F.G. Monteiro\altaffilmark{5}}
\affil{Instituto de Astrof\'{\i}sica e Ci\^encias do Espa\c{c}o, CAUP \& DFA/FCUP, Universidade do Porto, Rua das Estrelas, 4150-762 Porto, Portugal}

\begin{abstract}
We created a large database of physical parameters and variability indicators by fully reducing and analysing the large number
of spectra taken to complement the asteroseismic observations of the CoRoT satellite. 
7103 spectra of 261 stars obtained with the ESO echelle spectrograph HARPS have been stored in the VO-compliant database {\it Spectroscopic Indicators in a SeisMic Archive} (SISMA), along with the CoRoT photometric data of the 72 CoRoT asteroseismic targets. The remaining stars belong to the same variable classes of the CoRoT targets and were observed to better characterize the properties of such classes.
Several useful variability indicators (mean line profiles; indices of differential rotation, activity and emission lines) together with $v\sin i$ and radial velocity measurements have been extracted from the spectra. The atmospheric parameters T$_{\rm eff}, \log g$ and [Fe/H] have been computed following a homogeneous procedure.
As a result, we fully characterize a sample of new and known variable stars by computing several spectroscopic indicators, 
also providing some cases of simultaneous photometry and spectroscopy.
\end{abstract}


\section{Introduction}
The huge amount of stars observed from space needs for an intensive, time-consuming, ground-based follow-up to confirm 
both the correct identification of the source and its full characterization. This issue will rapidly grows in the future,
since new all-sky missions are ready to be launched or in advanced drawing. Therefore, it is of paramount importance 
to store the data and the information in a way that they can be exploited in the future to drive, complement, and help the
new analyses, making the incoming ground-based follow-ups more robust and straightforward.

In this context the ground-based activities performed in support of the {\it COnvection, ROtation and planetary Transits} \citep[CoRoT; ][]{corot} space mission are playing a relevant role.
The two main scientific programs of the CoRoT mission focused on asteroseismic studies and extrasolar planets search. Each program used two dedicated CCDs and had different observational strategies: the asteroseismic program observed up to 5 stars brighter than magnitude $m_{V}$=9 per CCDs (10 in total) per observing run, while the exoplanet program observed around 6000 stars per CCD (12000 in total) of magnitude between 11 and 16. CoRoT was launched on 2006 December 27 and it was retired on 2013 June 24 after a second, unrecoverable electronic failure occurred on 2012 November 2. The first failure, in 2009 March, caused the loss of two CCDs and the halving of the number of targets.

The whole asteroseismic program was flanked by a large ground-based observational effort ensuring a simultaneous monitoring \citep{corot-ground1, corot-ground4, corot-ground3, corot-ground2}.
Such a strategic decision was taken in the mission planning before launch in order to combine the high-quality photometry with simultaneous high-resolution spectroscopy.
The asteroseismic strategy of the space mission CoRoT was to monitor several different types of pulsating and variable stars, in order to cover almost the whole Hertzsprung-Russell diagram (Table~\ref{variable}). The observations were carried out on alternating short and long observing runs, of about 30 and 150 days respectively, focusing on up to 10 stars per run before the first failure and to 5 stars afterwards. 
\begin{table*}
\begin{center}
\caption{Type and number of variables in SISMA}
\begin{tabular}{crc}
\tableline
Variable type & \multicolumn{1}{c}{Label} &\multicolumn{1}{c}{Number of} \\
 & &\multicolumn{1}{c}{stars} \\
\tableline
\noalign{\smallskip}
\multicolumn{3}{c}{MAIN SEQUENCE STARS}\\
O-early B stars & OB & 8\\
Emission line B stars & Be & 41\\
Slowly pulsating B-type stars & SPB & 23\\
Slowly pulsating B-type stars-$\beta$ Cephei hybrid & SPB-bCep & 2\\
$\beta$ Cephei & bCep & 6\\
$\alpha^2$ Canum Venaticorum & a2CVn & 4\\
Chemically peculiar HgMn stars & HgMn & 2\\
late B-A-F stars & BAF & 13 \\
$\delta$ Scuti & dSct & 97\\
$\delta$ Scuti-$\gamma$ Doradus hybrid & dSct-gDor & 1\\
$\gamma$ Doradus & gDor & 22\\
T Tauri stars & TTau & 1\\
Solar-like stars & solar-like & 14\\
\noalign{\smallskip}
\noalign{\smallskip}
\multicolumn{3}{c}{EVOLVED STARS}\\
$\alpha$ Cygni & aCyg & 1\\
S Doradus & SDor & 2\\
Red giants & red-giant & 23\\
RR Lyrae & RRLyr & 1\\
\tableline
\label{variable}
\end{tabular}
\end{center}
\end{table*}

The ground-based activities started with the Large Programme 178.D-0361 using the FEROS spectrograph at the 2.2m telescope of the ESO-LaSilla Observatory, and continued with the Large Programmes LP182.D-0356 and LP185.D-0056 using the HARPS instrument at the 3.6m ESO telescope \citep{harps}. The spectroscopic survey branched out to several other high-resolution echelle spectrographs, i.e., SOPHIE at the Observatoire de Haute Provence, FOCES at Calar Alto Observatory, FIES and HERMES at the Roque de Los Muchachos Observatory, CORALIE at ESO-LaSilla Observatory, and HERCULES at the Mount John University Observatory.

The need to make such a large archive of high-resolution spectra available to the whole community became urgent with the end of the CoRoT mission. Therefore, a specific activity was included in the 
{\it SpaceInn: Exploitation of Space Data for Innovative Helio-and Asteroseismology}\footnote{\url{http://www.spaceinn.eu/}} project.
One of the purposes of this project is to establish and coordinate archives of helio- and asteroseismic data, both new and already existing. Furthermore, the project aims to make public the results of the analysis of these data and the tools that have been developed in treating them. As such, we pursued one of the {\it SpaceInn} goals by creating an archive to store not only the reduced spectra, but also a whole range of information on the variability content of the observed time series. In line with the project, we would like now to make the data accessible also to the broader scientific community beyond the CoRoT asteroseismic one.

Since HARPS was by far the most used instrument in our observational program, first we focused on the 7135 spectra collected on 261 targets. Next activities, perhaps in other projects, could be the full reduction of the more diversified spectra collected with the other spectrographs.
Thirty-two HARPS spectra have been discarded for quality reasons, and the remaining 7103 have been stored in the web-based VO-compliant archive {\it Spectroscopic Indicators in a SeisMic Archive} (SISMA), along with additional files and information. 
In this paper we illustrate the work done collecting the spectra, reducing them and obtaining the full set of spectroscopic indicators
(physical parameters, velocities, activity indices and so on) of stars belonging to several classes of variables.

\section{Observations and reduction of the HARPS spectra}
In the framework of the awarded two HARPS Large Programmes, 15 nights were allocated 
each semester over nine semesters, from 2008 December to 2013 January, for a total of 135 nights. 
The HARPS spectrograph covers the spectral range from 3780 to 6910~\AA, distributed over echelle orders
89-161.
We usually used it in the high-efficiency mode EGGS, with resolving power R=80,000
to obtain high signal-to-noise ratio (S/N) spectroscopic time-series.
Indeed, 
many of the CoRoT targets were hot stars, with large $v\sin{i}$ values, thus making the high-accuracy mode HAM (R=115,000) ineffective in improving the radial velocity precision.
We also note that the library of masks used by the online reduction pipeline of HARPS does not include hot star templates, again lowering the precision of the radial velocity measurements for hot stars, even in the case of low $v\sin{i}$ values. Moreover, the EGGS mode allowed us also to reduce the exposure times, and to increase the S/N, which is very useful for the line-profile variations (LPVs) analysis. Nevertheless, some specific targets (cool, bright stars with low $v\sin{i}$) have been observed in the high-accuracy mode HAM, more appropriate for obtaining very precise radial velocity measurements for these cases.
The statistical mode of the S/N of the EGGS spectra is S/N=220 at about 5800~\AA, 
while that of the HAM spectra is S/N=160 in the same region.

In addition to the primary targets, the satellite observed some specific asteroseismic targets in the exo CCDs, too. These objects are typically faint ($11\le m_{V}\le 16$) and unsuitable for detailed spectroscopic studies. Nevertheless, we took a few spectra of some of them, in order to obtain the physical parameters \citep[e.g., ][]{hads}. As requested in the case of observations conducted in Visitor Mode, other objects were observed as back-up and filling targets, aiming at better defining the physical properties of the variability classes observed by CoRoT.

All the HARPS spectra were reduced using a semi-automated pipeline developed at the Brera Observatory \citep{laurea}, which automatically normalize the spectra and retain information on the wavelength range of the orders (see Section~\ref{app:reduction} in the Appendix for more details on the reduction). The latter is very important for detailed spectroscopic analysis, because the S/N decreases greatly on the borders of the orders and the spectral lines in these regions may be distorted during the merging because of a lack of continuum on both sides of the lines. Our pipeline delivers two outputs for each spectrum: {\it a)} a five column table with wavelength, flux, normalized flux, S/N, and number of the echelle orders; {\it b)} a two-columns table with wavelength and normalized flux, with the echelle orders merged.
The normalization was done in an automated way, as such the normalized spectra are to be used with care, keeping in mind that the normalization could be unreliable in the first blue orders and on the borders of the orders.

We adopted a classification of the variable stars based on a preliminary analysis of the CoRoT light curves, HARPS spectra, and previous studies of the 261 targets (Table~\ref{variable}). In particular, several late B, A, and F stars selected as asteroseismic targets for CoRoT turned out to be constant or elusive variables. We grouped them under the ``BAF stars" label. On the other hand, the 
O and early-B stars class seems more intriguing, since the interplay between pulsation and rotation is still 
unclear in these hot stars \citep{blomme}. These targets, together with the calibration stars HD~34816 and HD~135240, are grouped under
the ``OB" label.

\section{Data analysis: indicators}
Our archive complements
the in-depth papers devoted to particular asteroseismic CoRoT targets \citep[e.g.,][]{50844,49330,50870,ana}.
For this purpose, we provide additional
information on the objects that can help in this endeavour by adding some useful indicators,
focusing on those that will help the study of stellar variability, caused either by pulsations, rotational activity or emissions.

\subsection{Mean line profiles: line-profile variations and rotational velocities} \label{sec:lsd}

We computed the mean line-profiles of each spectrum by means of the least-squares deconvolution (LSD) software \citep{lsd}. This method
greatly enhances the S/N, allowing to reach S/N$\gg$1000 \citep[e.g.,][]{50844,50870} since it grows roughly following $\sqrt{N}$, where
$N$ is the number of the used lines \citep[see Fig.~1 in ][]{lsd}.

\begin{figure}
\centering
\includegraphics[width=\linewidth]{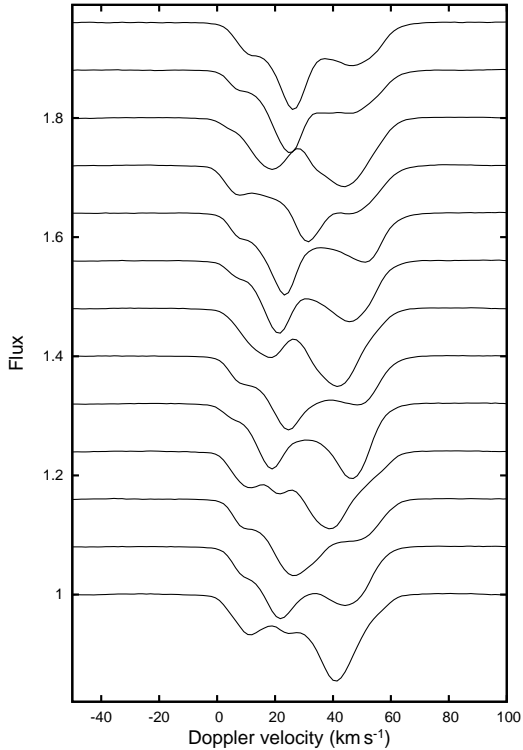}
\caption{Mean line profiles of the $\delta$~Scuti star HD~41641 at intervals of about 25~min. The fluxes have been shifted for easier interpretation. The line-profile variations are greatly enhanced by averaging a great number of spectral lines.}
\label{lpv}
\end{figure}

The LSD procedure requires as input normalized spectra and theoretical line masks (obtained from VALD, see~\citealt{vald}). The presence of saturated lines will greatly influence the shape of the resulting profiles, while telluric lines, which are not present in the mask, will increase the noise of the results. In order to obtain the best possible mean line profiles, we selected the
wavelength regions 4415-4805, 4915-5285, 5365-6505~\AA, i.e. we cut {\it a)} the blue end of the spectra (orders 139--161), where usually the S/N is very low and the automated normalization procedure may fail (see Section~\ref{app:reduction} in the Appendix), {\it b)} the red end of the spectra (orders 89-93), where most of the telluric lines are found, and {\it c)} the regions of the Balmer lines.
We used a 0.8~\kms\, step for the HAM spectra and a 1.4~\kms\, step for the EGGS spectra, aside for some cases where we were forced to compute the mean line profiles in a very large velocity range (up to 1000~\kms) to take into account very fast rotation in binaries, and as such we increased the step up to 4~\kms\, for computational reasons.
The mean line profiles are extremely useful to give clues of the importance of the nonradial oscillations. Figure~\ref{lpv} shows the clear perturbations due to a nonradial pressure mode in the mean line--profiles of the $\delta$ Sct star HD~41641: the most probable identification of such a mode is $(\ell,m)=(3,1)$ \citep{ana}. The line-profile variations also provide the most stringent piece of evidence that high-degree $\ell$ modes ($\ell\ge 6$) are excited in $\delta$ Sct stars (see Fig.~8 in \citealt{50844} and Fig.~13 in \citealt{50870}).

\begin{figure}
\centering
\includegraphics[width=\linewidth]{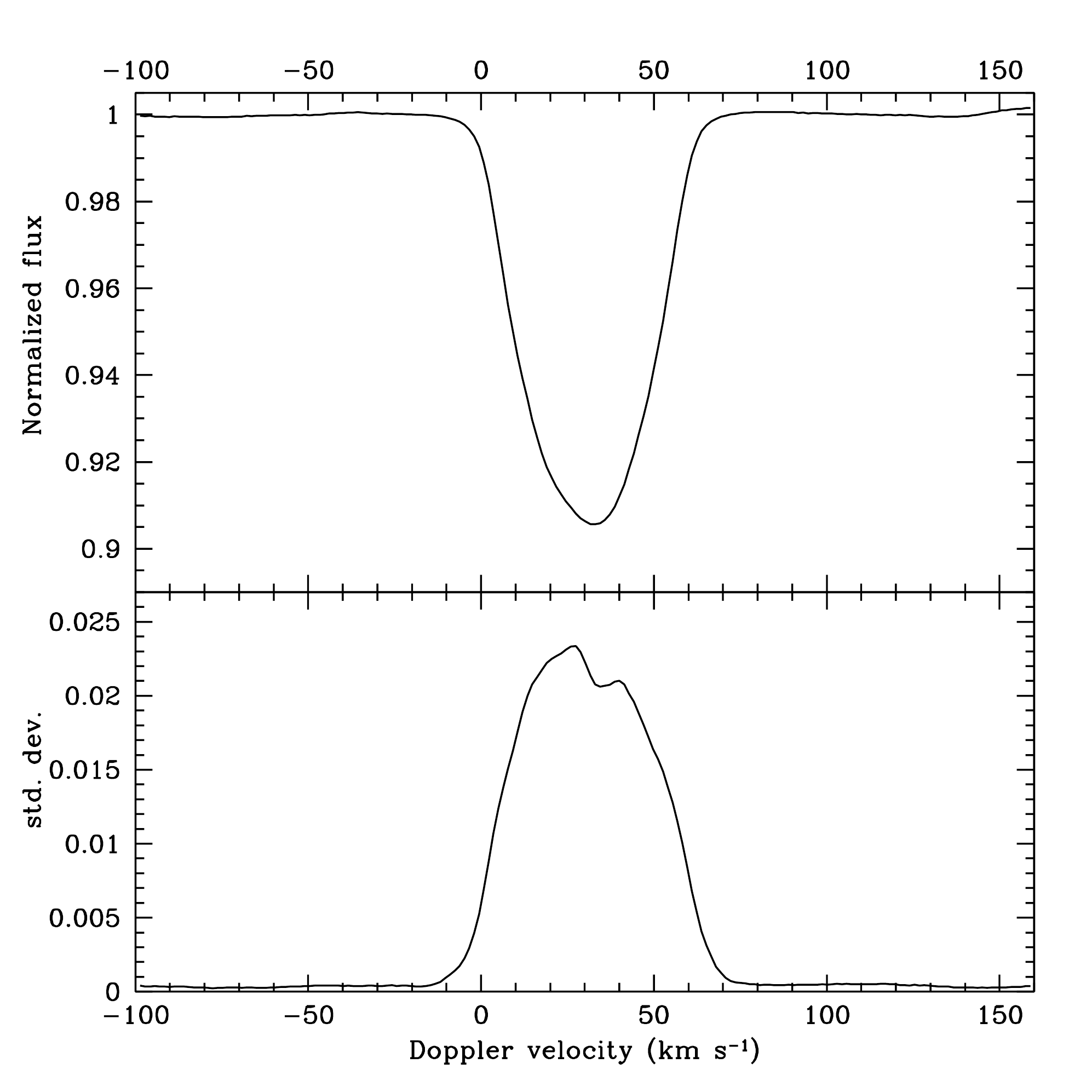}
\caption{Average mean line profile and standard deviations for the $\delta$~Scuti star HD~41641. The line-profile variations are clearly concentrated at the center of the line.}
\label{profmed02}
\end{figure}

\begin{figure}
\centering
\includegraphics[width=\linewidth]{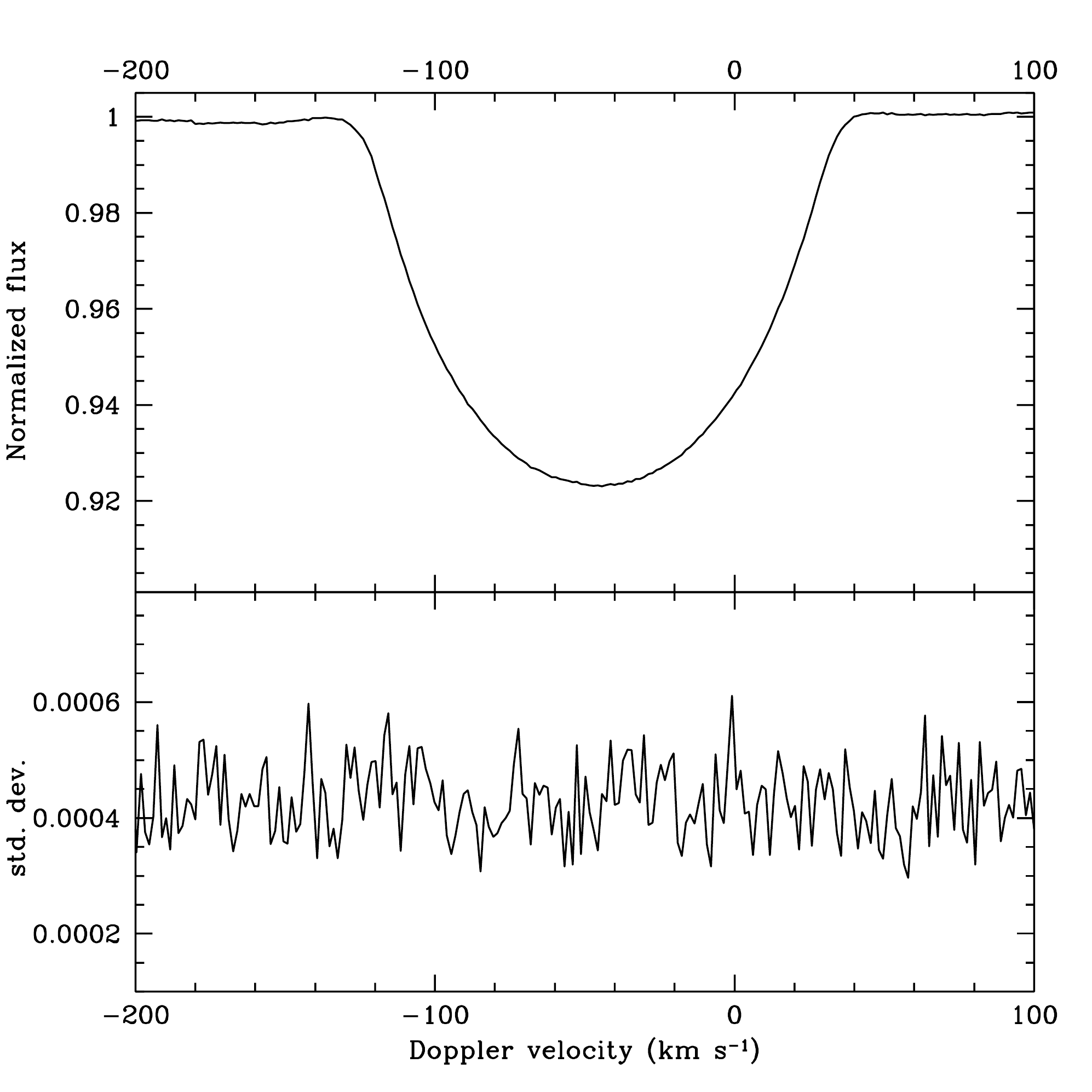}
\caption{Average mean line profile and standard deviations for A-type star HD~170133, 
where no line-profile variations are present.}
\label{profmed01}
\end{figure}

In order to provide a quick look at the pulsational content of the time series in the archive, we created for each target a PDF figure with the average of the mean line profiles of the spectra and their standard deviation from the average. Stars exhibiting line-profile variations clearly show a variable standard deviation along the average line profile, giving also hints where the deviations are the largest ones (see Fig.~\ref{profmed02} again for the case of HD~41641). On the contrary, time series without line-profile variations show a constant standard deviation. Figure~\ref{profmed01} illustrates the clear constant behaviour of the line profiles of HD~170133, an A-type star found to be constant also in the CoRoT photometric time series.

We used the mean line profiles not only to measure the $v\sin{i}$ of the stars but, if possible, to indicate the presence of differential rotation, too. This is done using the Fourier transform of the symmetrized mean line profiles, obtained by mirroring the profiles at the center and averaging the original and mirrored profiles.
The position of the first zero of the transform, $q_{1}$, is used to compute the $v\sin{i}$ value. The $q_{2}/q_{1}$ ratio of the first two zero positions is an indication of possible solar-like differential rotation (if $q_{2}/q_{1}<1.72$), anti-solar differential rotation (if $q_{2}/q_{1}>1.83$), or rigid rotation \citep[if $1.72<q_{2}/q_{1}<1.83$;][]{reiners}. 
This only works in case of rotators with $v\sin{i}>10~$~\kms, where the rotation is the main contributor to the line broadening; otherwise we estimated the $v\sin{i}$ from the FWHM of the gaussian fitting of the mean line profiles and set the $q_{2}/q_{1}$ indicator as null.

The region of uncertainty around the zero positions of the Fourier transform is given by $S_{f} = S_{p} \Delta x N^{1/2}$, where $S_{p}$ is the standard deviation between the mean line profile and its mirrored profile, $\Delta x$ is the step of the data and $N$ is the number of data in the line profile. As such, the errors increase with the broadening of the lines and when the profiles are heavily deformed by pulsations. Then we propagated the errors linearly to find the uncertainties on $v\sin{i}$ and $q_{2}/q_{1}$.

\begin{figure}
\centering
\includegraphics[width=\linewidth]{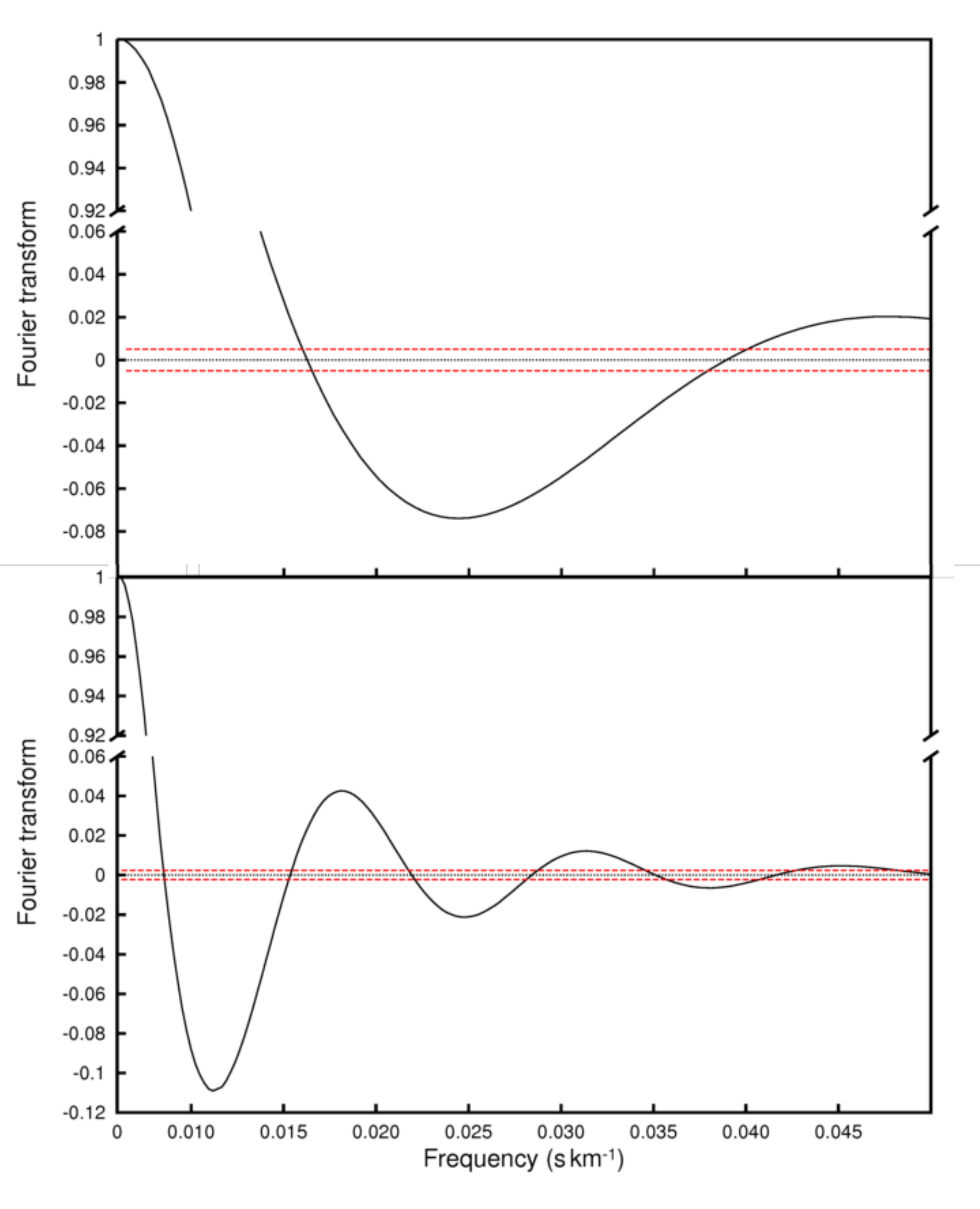}
\caption{{\it Top panel:} Fourier transform of the mean line profile of a spectrum of the $\delta$ Sct star HD~41641 (S/N=173). {\it Bottom panel:} The same for a spectrum of the non-variable star HD~170133 (S/N=152). The red dashed lines show the regions of uncertainty around the zero positions.}
\label{fourier}
\end{figure}

As an example, we obtained $v\sin{i}=77.3\pm0.2$~\kms\, and $q_{2}/q_{1}=1.80\pm0.01$ for a spectrum of the non-variable star HD~170133 with S/N=152 (Fig.~\ref{fourier}, bottom panel). The uncertainty on the zero position is very small due to the excellent mean line profile (Fig.~\ref{profmed01}).
The situation is more complicated for the $\delta$ Sct variable HD~41641. Examining a spectrum with S/N=173, the effects of the pulsations on the mean line profiles resulted in much larger errors, i.e. $v\sin{i}=29.8\pm2.2$~\kms\, and $q_{2}/q_{1}=1.76\pm0.26$ (Fig.~\ref{fourier}, top panel). The indicators of differential rotation suggest rigid-body models for both stars.

We provided the radial velocities computed by the HARPS pipeline for the HAM spectra, while we computed those of the EGGS spectra by fitting the mean line profiles with a Gaussian. Indeed, it was not possible to use the simultaneous lamp calibration in the EGGS mode because the dedicated fiber is broken. Moreover, most of our EGGS targets are hotter than the masks used by the ESO pipeline to compute the cross-correlation function: as such, the radial velocity computed by the HARPS pipeline is not as accurate as in the HAM mode.

\subsection{Activity index}
Our index is not calibrated on the Mount Wilson one, since it is intended to be used to study variations only, not absolute values of activity.
We decided to provide an indication of the stellar activity in order to better characterize the targets and to help study the stellar variations due to the pulsations along with activity variations. Because of the large range of spectral types, we defined a simple index using the H and K lines of Ca~II \citep{rainer}.
To do this an automated procedure shifts the spectra by their radial-velocity values and applies a specific normalization around the H and K lines.
The latter step is necessary because our pipeline performs a global normalization on the whole spectrum for all different kinds of stars, and as such it may underestimate the continuum in the blue orders, where the S/N is usually very low especially in the case of cool stars (see Section~\ref{app:reduction} in the Appendix). A simple linear interpolation using two continuum regions on the two sides of the Ca~II spectral lines (3987-3918~{\AA} and 4000-4080~{\AA}) is enough to correct this problem.
Then the program computes the areas of four spectral regions: two regions of 2.5~{\AA} each in the center of the H and K lines ($A_{H}$ and $A_{K}$), and two regions of 10~{\AA} each on the continuum ($A_{c1}$ and $A_{c2}$).

We obtained two indices, one for the H line and one for the K line:
\begin{equation}
I_{H} = \frac{A_{H}}{(A_{c1} + A_{c2})/2}
~~~~~~~~I_{K} = \frac{A_{K}}{(A_{c1} + A_{c2})/2}
\end{equation}
The final index is simply the average of the two individual ones:
\begin{equation}
I_{HK} = \frac{I_{H} + I_{K}}{2}=\frac{A_{H}+A_{K}}{A_{c1} + A_{c2}}
\end{equation}

Values of $I_{HK}$ larger than 0.2--0.3 indicate the presence of activity in the specific HARPS spectrum. The index is very useful to trace activity of solar-like stars, less for hotter star (spectral type A or earlier), but it is automatically computed and as such it is given for all the spectra. The activity index can be relied upon for more than one hundred stars of our sample.

\subsection{Violet-to-red peak intensity ratio in Be stars}
There are 40 stars in our archive that show emission lines. Mostly they are Be stars, i.e. stars from late O to early A spectral types surrounded by circumstellar disk fed by stellar mass loss outburts: the emission lines of these stars are produced both by the stellar outbursts and by the disk.

\begin{figure}
\centering
\includegraphics[width=\linewidth]{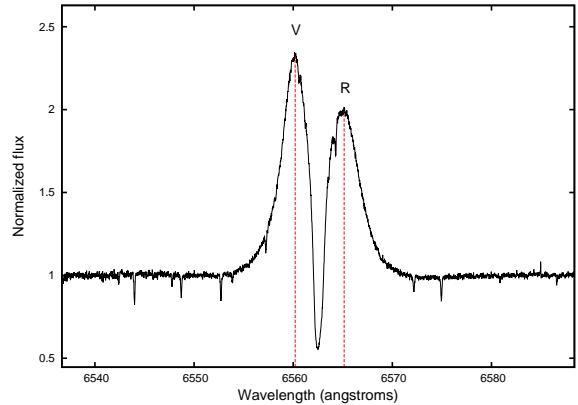}
\caption{Observed spectrum of the Be star HD~171219 in the H$\alpha$~region. The violet and red peaks of the emission are shown with dashed lines.}
\label{havr}
\end{figure}

Be stars show spectral and photometric variability on wide timescales, ranging from minutes to years; a interesting quantity to study is the violet-to-red peak-intensity ratio (V/R) which is the ratio between the peak intensity of the violet and the red part of the H$\alpha$ emission, and as such it is an indicator of the line asymmetry \citep{saad}. Figure~\ref{havr} shows the H$\alpha$ profile of the Be star HD~171219 (Andrade et al., in prep.).
In most cases, Be stars have stable V/R equal to 1 (no asymmetry), but roughly a third of them show instead quasi-periodic V/R variations, which may be a combination of orbital period and long term variability (on a timescale of several years).
 
All the spectra in our archive have a true/false flag for the presence of emission in the H$\alpha$ line. In the case of emission, the value of V/R is given; if there is no emission, V/R is set to zero.

\section{Data analysis: physical parameters}
The physical parameters of all the stars in the database are estimated with the spectral synthesis method using the 3.3 version of the SME software \citep{sme} and the stellar line lists from VALD~\citep{vald} on a selected wavelength region (5160-5190~\AA). We usually have several spectra for each star: in these cases we shifted the spectra by their measured radial velocities and then we averaged them, in order to lessen the effect of the pulsations on the shape of the stellar lines. Then we computed the physical parameters ($\rm{T_{eff}}$, $\log{g}$, and [Fe/H]) on the average spectrum.

To estimate their error bars we calculated the external uncertainties, i.e.,
the differences between the SISMA values and those reported in the
papers where the
detailed analyses of the CoRoT targets were performed. We consider these external uncertainties more realistic
than the standard deviations of the parameters of the same star obtained from different
spectra (i.e., internal uncertainties).
In evaluating the external uncertainties we have to keep in mind that both determinations
are affected by their own errorbars and we are comparing a single method to several different case-by-case ones.
We could use a sample of 25 stars.
We obtained mean differences of 240~K (i.e., 4.0\%) in $\rm{T_{eff}}$, 0.32~dex
in $\log{g}$, and 0.33~dex in [Fe/H] for cool ($\rm{T_{eff}}<$~6000~K)
main-sequence stars.
These values are very close to the external uncertainties obtained from
the LAMOST parameters and the literature ones: 3.5\%, 0.3~dex, and 0.2~dex \citep{lamost}.
The uncertainties increase to
400~K in $\rm{T_{eff}}$, 0.50~dex in $\log{g}$, and 0.40~dex in [Fe/H]
when extending the sample to early-type stars,
as expected by including very fast rotators, such as B-type stars. Indeed, their
parameters are strongly method-dependent due to line blending and their almost complete
dilution in the continuum.

\begin{figure}
\centering
\includegraphics[width=\linewidth]{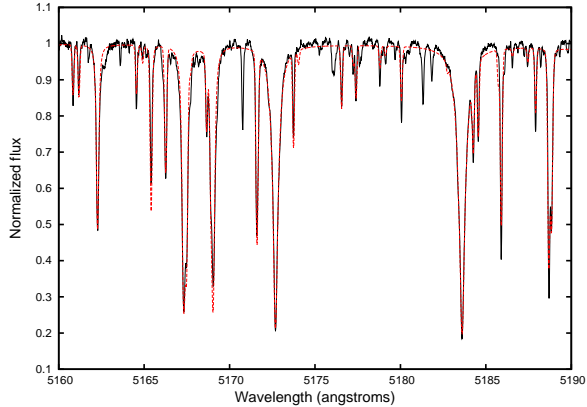}
\caption{Observed spectrum of the $\delta$~Scuti star HD~39996 (solid black line) and best synthetic fit (red dashed line).}
\label{sme-fit}
\end{figure}

\begin{figure}
\centering
\includegraphics[width=\linewidth]{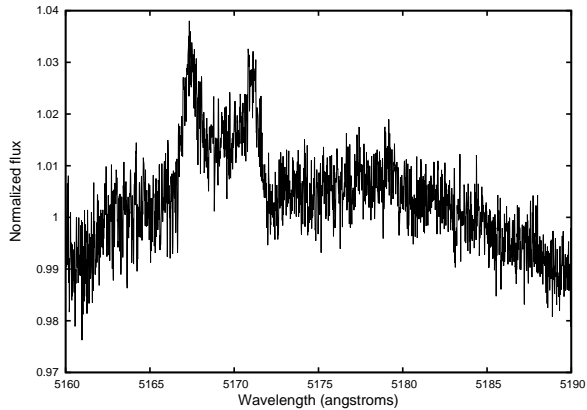}
\caption{Observed spectrum of the Be star HD~128293 in the 5160-5190~{\AA}~region. The emissions are clearly visible. It is not possible to fit this spectrum in order to estimate the physical parameters of the star.}
\label{sme-nofit}
\end{figure}

To allow the user to have an immediate feeling of the goodness of the given set of the parameters, we also provide for each star a PDF figure with the observed spectrum and the best fit (Fig.~\ref{sme-fit}).
Under very special circumstances (e.g., very fast rotators, emission line stars, binary stars, and so on), we were not able to estimate the physical parameters with this automated method, as for example in the specific case of the strong emission lines of the Be star HD~128293 (Fig.~\ref{sme-nofit}). In these cases, a PDF figure with only the observed spectrum in the 5160-5190 {\AA}~region is given in the archive.

Finally, we made available online a table with the estimated physical parameters and other information (variability type, binarity, and so on). A portion is shown in Tab.~\ref{parameters}. 

\begin{table*}
\begin{center}
\caption{Table~\ref{parameters} is available online. A portion is shown here for guidance
regarding its form and content.}
\begin{tabular}{lrrrrrrrr}
\tableline
\multicolumn{1}{c}{Target name} & \multicolumn{1}{c}{Number of}& \multicolumn{1}{c}{$\rm{T_{eff}}$} & \multicolumn{1}{c}{$\log{g}$} & \multicolumn{1}{c}{[Fe/H]} & \multicolumn{1}{c}{Binary} & \multicolumn{1}{c}{Variable} & \multicolumn{1}{c}{H$\alpha$} & \multicolumn{1}{c}{S/N} \\
 & \multicolumn{1}{c}{spectra} & \multicolumn{1}{c}{[K]} & \multicolumn{1}{c}{[cgs]} & & \multicolumn{1}{c}{flag} & \multicolumn{1}{c}{type} & \multicolumn{1}{c}{emission} & \multicolumn{1}{c}{range}\\
\tableline
BD184914 & 1 & 7162 & 3.21 & 0.43 & no & dSct & no & 93\\
GSC00144-03031 & 3 & 7822 & 3.41 & 0.09 & no & dSct & no & 109-121\\
HD001097 & 1 & 6595 & 3.70 & 0.31 & no & dSct & no & 123\\
HD007312 & 1 & 8378 & 3.88 & 0.55 & no & dSct & no & 216\\
HD008781 & 1 & 7280 & 2.91 & -0.23 & no & dSct & no & 132\\
HD009065 & 35 & 7471 & 3.30 & -0.41 & no & dSct & no & 129-378\\
HD009133 & 1 & 8812 & 4.25 & 0.49 & no & dSct & no & 117\\
HD010167 & 1 & ... & ... & ... & SB2 & gDor & no & 265\\
HD011462 & 10 & 11891 & 4.05 & 0.07 & no & bCep & no & 74-194\\
HD011956 & 1 & 8986 & 3.93 & 0.71 & no & dSct & no & 197\\
HD016031 & 14 & 5633 & 4.38 & 0.13 & no & no & no & 57-89\\
HD016189 & 65 & 7338 & 3.72 & 0.00 & no & dSct & no & 131-419\\
... & ... & ... & ... & ... & ... & ... & ... & \\
\tableline
\label{parameters}
\end{tabular}
\end{center}
\end{table*}

\section{CoRoT photometry}
The most observed stars, i.e. those with more spectra in the archive, are the main asteroseismic targets of the CoRoT Long Runs. In fact, the ground-based campaigns were planned to have the spectroscopic counterpart of the CoRoT light curve. The light curves of these 72 stars have been retrieved from the CoRoT public archive and stored in our database alongside the spectra. In some cases, the targets were observed with CoRoT in more than a Long Run resulting in more than one set of photometric data per target.
The CoRoT data are the Legacy light curves \citep{corotbook}\footnote{\url{
http://open.edpsciences.net/index.php?option=com_content&view=article&id=317}}.

The light curves are provided as FITS files, with some auxiliary information stored in the header. 
The FITS files contain three tables: RAW (N1 data with correction from aliasing, offsets, background and satellite's jitter), BAR (N2 data with some flux correction for thermal effects and long term efficiency, and spurious points replaced by interpolation), and 
BARREG \citep[BAR N2 data with invalid and missing data replaced using the Inpainting method;][]{pires}.

\begin{figure}
\centering
\includegraphics[width=\linewidth]{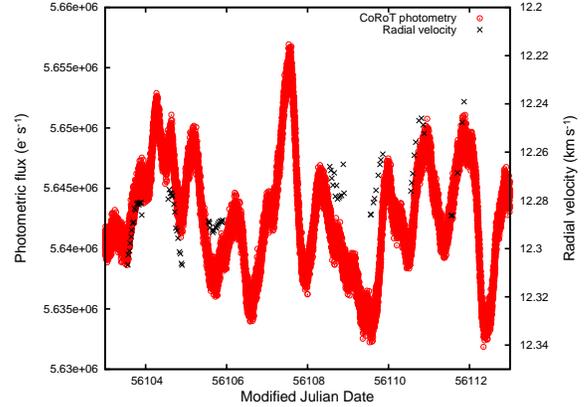}
\caption{Zoom of the simultaneous plot of the CoRoT light curve (red dots) and the radial velocity curve (black crosses) of the red giant HD~178484. Instead of the radial velocities, it is possible to plot any of the other spectroscopic indicators, {\textit{i.e.}} $v\sin{i}$, $q_{2}/q_{1}$, activity indices or V/R values.}
\label{archivio03}
\end{figure}

The nights with HARPS were scheduled during the period of best visibility of the CoRoT targets and usually we were able to get the requested simultaneous spectroscopy and photometry. However, the time baseline of the spectroscopic observations is much smaller than that of the photometric ones (typically 20-25 nights against 30-150 days).
Nevertheless, it was possible to monitor the spectroscopic counterpart of the photometric variations: Figure~\ref{archivio03} shows an example of simultaneous radial velocity and flux measurements.

\section {Conclusions}
With the large number of space missions and ground-based campaigns that are continuously carried on, it becomes important to store the large amount of observational data in an easily accessible database, so that it can be exploited by the whole scientific community, both to re-analyse the available data with different tools and to help in the preparation of future missions and related archives.

The ground-based spectra played a key role in the exploitation of the CoRoT mission thanks to the coordinated starting of the spectroscopic and photometric campaigns.
Towards this goal, we built the online database SISMA using the high-resolution HARPS spectra observed simultaneously with the CoRoT satellite, and the relative light curves. 
While the CoRoT targets, with their simultaneous spectroscopy and photometry, are the backbone of the archive, we would like to highlight the importance of the many non-CoRoT targets, aimed at improving the study of classes of variable stars by means of the provided indicators, for the most part still unpublished and unused in a statistical sense.
The high S/N, high-resolution, fully reduced and calibrated archival spectra are enriched with many observational parameters to be directly used as tools to investigate the spectroscopical effects of the stellar variability: mean line profiles, differential rotation indicators, activity indices, H$\alpha$ violet-to-red peak-intensity ratio along with the computed $v\sin{i}$ and radial velocity values.
In addition to that, the stellar atmospherical parameters available in the archive are computed in an homogeneous way.
The structure of the archive is explained in the Section~\ref{app:archive} of the Appendix.

The SISMA archive offers the possibility to study the stellar variability from different perspectives. To give a few examples,
{\it i)} the interplay between rotation rate and distribution of pulsational frequencies in hot variables close to the Main Sequence. The link between rotational frequencies, large separations, and regular spacings is a promising topic in the study of $\delta$ Sct stars \citep{garcia,paparo} and $\gamma$~Dor stars \citep{van};
{\it ii)} the presence of circumstellar, narrow emission structures in the core of absoption lines of fast rotators indicating the rate occurrency of mass-loss events;
{\it iii)} the relation between activity indices and differential rotation in cool stars, a very useful tool to investigate the star-planet interaction \citep{borsa}.

We feel that the SISMA database fully matches the need to provide a set of 
high-quality data, to be used beyond the asteroseismic community that planned the observations and for purposes much wider that
those driving them at the beginning of the CoRoT adventure.

\acknowledgments
This work is based on observations made with the 3.6-m telescope at La Silla Observatory under the ESO Large Programmes LP182.D-0356 and LP185.D-0056.
The research leading to this result has received funding from the European Community's Seventh Framework Programme ([FP7/2007-2013]) under grant agreement no. 312844. The authors thank Artie Hatzes for useful comments on a first draft of
the manuscript.

\facilities{ESO:3.6m (HARPS spectrograph), CoRoT}
\software{LSD \citep{lsd}, 
          SME \citep{sme}, 
          VALD \citep{vald}, 
          HARPS DRS, RGraph, STIL}

\appendix

\section{The reduction of the HARPS spectra} \label{app:reduction}
The HARPS Data Reduction Software pipeline (DRS) automatically reduces all the observed spectra by doing bias subtraction, masking of bad columns, dark subtraction, flat-fielding, extraction of the orders, removal of the blaze function, wavelength calibration, merging of the orders and radial velocity computation. The final results are the merged monodimensional *s1d* files, but we were dissatisfied with the blaze removal and we felt that the merging procedure resulted in loss of information on the quality of some spectral regions.

Our reduction pipeline worked with the *e2ds* FITS files given by the online DRS, i.e. the bi-dimensional spectra with the echelle orders extracted and corrected for the flat-field. These first steps are done quite well by the online pipeline, so there was no need to re-do them. Because one of our goals was to easier normalize the spectra, we took particular care in removing the blaze function from the orders. In order to do this, we observed every night a very bright hot star (HD~34816, spectral type B0.5, or HD~135240, spectral type O8, according to their visibility): 
because of their few spectral lines, we could easily fit their continuum and use it as the blaze function.

We also computed the S/N ratio of the spectra for each pixel by using the expression:
\begin{equation}
S/N = \frac{1}{\sqrt{ {{\epsilon}_{spectrum}}^{2} + {{\epsilon}_{flat}}^{2} }}
\end{equation} 
where ${\epsilon}_{spectrum,flat}$ are given by:
\begin{equation}
{\epsilon}_{spectrum,flat} = \frac{\sqrt{RON^{2} + gain \times N_{spectrum,flat}}}{gain \times N_{spectrum,flat}}
\end{equation}
where the read-out-noise $RON$ is $4.5~e^{-}$, the gain is $0.75~ADU/e^{-}$ and $N_{spectrum,flat}$ are the counts per pixel in the science and flat-field spectra, respectively. This allows us 
to study the spectra in detail keeping in mind the reliability of each part of the spectra.

Then we divided the spectra for the blaze function, calibrated them in wavelength, corrected them for their barycentric velocity and rebinned them with a 0.1~\AA\, step. 
Afterwards we normalized the spectra in an automated way. Our pipeline normalized each order with an iterative procedure using a second-degree polynomial computed from a fixed point ${\lambda}_{0}$ chosen in the middle of the order, far from the borders:
\begin{itemize}
\item{our pipeline computed a first polynomial fit;}
\item{our pipeline computed the residuals between the spectrum and the polynomial, all the pixels with a residual larger than $3\sigma$ are considered spectral lines, along with their adjacent pixels until the sign of the residuals changes;}
\item{our pipeline computed another fit on the spectrum minus the spectral lines;}
\item{our procedure iterated until it converged.}
\end{itemize}
Some orders may have a lot of spectral lines, in which case the normalization could be uncorrect. To remedy this situation, our pipeline confronted the polynomials coefficients found for all the orders, which should be very similar in case of a correct normalization. If they differed a lot (which happened only for a couple of orders at worst), the uncorrect polynomial coefficients were substituted by fitting the coefficients found for all the spectral orders. Unfortunately several adjacent blue orders of the spectra have this problem, worsened by the lower S/N ratio in these regions due to the efficiency of the CCD, that translates in a underestimation of the continuum in the first blue orders of the spectra.

The merging of the orders was done on the overlapping continuum regions of the normalized spectra. The pixels are weighted using their S/N ratio and averaged. As such our procedure outputted two-columns normalized merged spectra (wavelength and normalized flux) and five-columns normalized and un-normalized unmerged spectra (wavelength, flux, normalized flux, S/N, and echelle orders).

\section{The ensemble of the stellar parameters and the structure of the archive} \label{app:archive}
All the data (reduced spectra, indicators and photometric series) are stored as either FITS or PDF files in the SISMA archive and can be accessed at \url{http://sisma.brera.inaf.it/}. The data can also be accessed through the Seismic Plus portal\footnote{\url{http://voparis-spaceinn.obspm.fr/seismic-plus/}}, developed in the framework of the {\it SpaceInn} project in order to gather and help coordinated access to several different solar and stellar seismic data sources.

\subsection{Database infrastructure}
The operating system for the online archive (database and web pages) is Linux assembled under the model of free and open-source software development and distribution. The object-relational database management system (ORDBMS) is PostgreSQL, a free and open-source software developed by the PostgreSQL Global Development Group. The ORDBMS manages all data and files stored in the archive.

The web user interface to the database is programmed using the Java/Servlet/JSP language, that creates web-based applications using Model View Controller. The web pages are Java Server Pages (JSPs) with embedded Java, HTML and CSS codes. The codes are executed on the server, then the page is returned to the browser for display. The server for Java Servlet and JSPs technologies is Apache Tomcat.

JSPs use RGraph\footnote{\url{http://www.rgraph.net/}} for charts visualization, a JavaScript charts library that produces the charts dynamically with JavaScript and the HTML canvas tag. The charts are made inside the browser with JavaScript so they are both quick to create and small. The archive can be queried either by target (using the name resolver service Sesame\footnote{\url{http://cdsweb.u-strasbg.fr/doc/sesame.htx}}) or by variability class, and the data can be retrieved or displayed.
In the latter case, it is possibile to plot the spectra, the light curves and the spectroscopic time series online. The library used to read the FITS files is STIL (Starlink Tables Infrastructure Library)\footnote{\url{http://www.star.bristol.ac.uk/~mbt/stil/}}.

\subsection{Scientific data}
The database contains several files for each observed object:
\begin{itemize}
\item{$\ast$\_full.fits: the main deliverables, i.e., the reduced spectra. The five columns list wavelength (barycentric corrected, in~\AA), reduced flux, normalized flux, signal-to-noise ratio, and echelle order;}
\item{$\ast$\_nor.fits: additional reduced spectra, automatically normalized and with the orders merged. They have two columns (barycentric corrected wavelength in~\AA, and normalized flux). It is important to keep in mind that the normalization is done by an automated procedure and as such a careful check is necessary for a detailed study of any particular line;}
\item{$\ast$\_mean.fits: the mean line profiles of each spectrum computed with the LSD software in the 4415-4805, 4915-5285, 5365-6505~{\AA} wavelength ranges. The files consist of two columns: Doppler velocity and normalized flux;}
\item{OBJECT\_tbl.fits: a general overview of the object time series:
\begin{itemize}
\item{the root names of the spectra,}
\item{the barycentric Julian dates at mid-exposure,}
\item{the signal-to-noise ratios of the spectra in the 5805-5825~{\AA}~region,}
\item{the radial velocities of the spectra and their errors,}
\item{the projected rotational velocities $v\sin{i}$ of the spectra and their errors,}
\item{the $q_{2}/q_{1}$ values and their errors,}
\item{the activity index $I_{H}$, using only the Ca~II H line, of each spectrum,}
\item{the activity index $I_{K}$, using only the Ca~II K line, of each spectrum,}
\item{the averaged activity index $I_{HK}$ of each spectrum,}
\item{the H$\alpha$ violet-to-red peak-intensity ratio (V/R) of each spectrum,}
\item{in the case of double or multiple systems, the radial velocities, v$\sin{i}$ and their errors will be listed for all the components, if possible;}
\end{itemize}}
\item{OBJECT\_profmed.pdf: a PDF file that allows a quick look at the line-profile variations of the time series.
In the case of objects where a single spectrum was observed, the mean line profile of the spectrum is given instead;}
\item{OBJECT\_fit.pdf: a PDF file with the observed spectrum in the 5160-5190~{\AA}~region and the best-fit synthetic spectrum.
In the case of objects where the fit was not possible, only the observed spectrum is given.}
\item{CoRoT light curves: FITS files with the CoRoT photometry. The light curves are available only for the 72 CoRoT targets having
at least one HARPS spectrum. }
\end{itemize}

To summarize, SISMA contains reduced photometric and spectroscopic observations, and time series of spectroscopic indicators. The physical parameters of the stars and the various indices are properly stored also in the FITS headers of the reduced spectra.

\end{document}